\documentclass[twoside,twocolumn,9pt]{article}
\usepackage{extsizes}
\usepackage[super,sort&compress,comma]{natbib} 
\usepackage[version=3]{mhchem}
\usepackage[left=1.5cm, right=1.5cm, top=1.785cm, bottom=2.0cm]{geometry}
\usepackage{balance}
\usepackage{mathptmx}
\usepackage{sectsty}
\usepackage{graphicx} 
\usepackage{lastpage}
\usepackage[format=plain,justification=justified,singlelinecheck=false,font={stretch=1.125,small,sf},labelfont=bf,labelsep=space]{caption}
\usepackage{float}
\usepackage{fancyhdr}
\usepackage{fnpos}
\usepackage[english]{babel}
\addto{\captionsenglish}{%
  
}
\usepackage{array}
\usepackage{droidsans}
\usepackage{charter}
\usepackage[T1]{fontenc}
\usepackage[usenames,dvipsnames]{xcolor}
\usepackage{setspace}
\usepackage[compact]{titlesec}
\usepackage{hyperref}

\usepackage{epstopdf}

\definecolor{cream}{RGB}{222,217,201}

\begin{document}

\pagestyle{fancy}
\thispagestyle{plain}
\fancypagestyle{plain}{
\renewcommand{\headrulewidth}{0pt}
}

\makeFNbottom
\makeatletter
\renewcommand\LARGE{\@setfontsize\LARGE{15pt}{17}}
\renewcommand\Large{\@setfontsize\Large{12pt}{14}}
\renewcommand\large{\@setfontsize\large{10pt}{12}}
\renewcommand\footnotesize{\@setfontsize\footnotesize{7pt}{10}}
\makeatother

\renewcommand{\thefootnote}{\fnsymbol{footnote}}
\renewcommand\footnoterule{\vspace*{1pt}%
\color{cream}\hrule width 3.5in height 0.4pt \color{black}\vspace*{5pt}} 
\setcounter{secnumdepth}{5}

\makeatletter 
\renewcommand\@biblabel[1]{#1}            
\renewcommand\@makefntext[1]%
{\noindent\makebox[0pt][r]{\@thefnmark\,}#1}
\makeatother 
\renewcommand{\figurename}{\small{Fig.}~}
\sectionfont{\sffamily\Large}
\subsectionfont{\normalsize}
\subsubsectionfont{\bf}
\setstretch{1.125} 
\setlength{\skip\footins}{0.8cm}
\setlength{\footnotesep}{0.25cm}
\setlength{\jot}{10pt}
\titlespacing*{\section}{0pt}{4pt}{4pt}
\titlespacing*{\subsection}{0pt}{15pt}{1pt}

\fancyfoot{}
\fancyfoot[LO,RE]{\vspace{-7.1pt}\includegraphics[height=9pt]{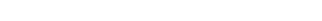}}
\fancyfoot[CO]{\vspace{-7.1pt}\hspace{11.9cm}\includegraphics{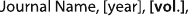}}
\fancyfoot[CE]{\vspace{-7.2pt}\hspace{-13.2cm}\includegraphics{head_foot/RF}}
\fancyfoot[RO]{\footnotesize{\sffamily{1--\pageref{LastPage} ~\textbar  \hspace{2pt}\thepage}}}
\fancyfoot[LE]{\footnotesize{\sffamily{\thepage~\textbar\hspace{4.65cm} 1--\pageref{LastPage}}}}
\fancyhead{}
\renewcommand{\headrulewidth}{0pt} 
\renewcommand{\footrulewidth}{0pt}
\setlength{\arrayrulewidth}{1pt}
\setlength{\columnsep}{6.5mm}
\setlength\bibsep{1pt}

\makeatletter 
\newlength{\figrulesep} 
\setlength{\figrulesep}{0.5\textfloatsep} 

\newcommand{\topfigrule}{\vspace*{-1pt}%
\noindent{\color{cream}\rule[-\figrulesep]{\columnwidth}{1.5pt}} }

\newcommand{\botfigrule}{\vspace*{-2pt}%
\noindent{\color{cream}\rule[\figrulesep]{\columnwidth}{1.5pt}} }

\newcommand{\dblfigrule}{\vspace*{-1pt}%
\noindent{\color{cream}\rule[-\figrulesep]{\textwidth}{1.5pt}} }

\makeatother

\twocolumn[
  \begin{@twocolumnfalse}
{
{\includegraphics[height=55pt]{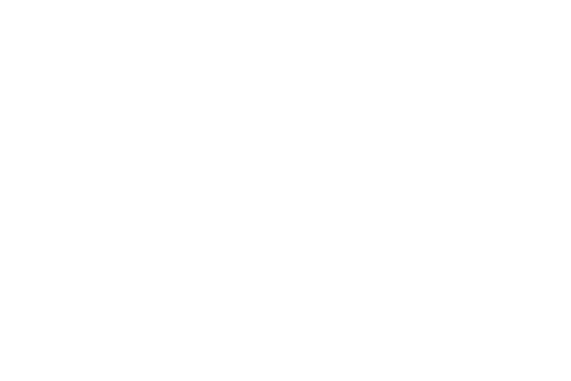}}\\[1ex]
\includegraphics[width=18.5cm]{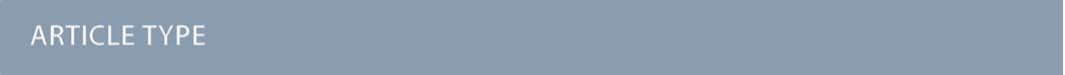}}\par
\vspace{1em}
\sffamily
\begin{tabular}{m{4.5cm} p{13.5cm} }

& \noindent\LARGE{\textbf{Surface hopping molecular dynamics simulation of ultrafast methyl iodide photodissociation mapped by Coulomb explosion imaging}} \\
\vspace{0.3cm} & \vspace{0.3cm} \\

 & \noindent\large{Yijue Ding$^{\ast}$\textit{$^{a}$}, Loren Greenman\textit{$^{a}$} and Daniel Rolles\textit{$^{a}$}} \\
 

\includegraphics{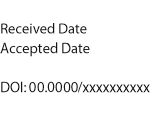}
& \noindent\normalsize{We present a highly efficient approach to directly and reliably simulate the photodissociation followed by Coulomb explosion of methyl iodide. In order to achieve statistical reliability, more than 40,000 trajectories are calculated on accurate potential energy surfaces of both the neutral molecule and the doubly charged cation. Non-adiabatic effects during photodissociation are treated using a Landau-Zener surface hopping algorithm. The simulation is performed analogous to a recent pump-probe experiment  using coincident ion momentum imaging
[Ziaee \textit{et al., Phys. Chem. Chem. Phys.} 2023, \textbf{25}, 9999]. 
At large pump-probe delays, the simulated delay-dependent kinetic energy release signals show overall good agreement with the experiment, with two major dissociation channels leading to $\text{I}(^2\text{P}_{3/2})$ and $\text{I}^*(^2\text{P}_{1/2})$ products. At short pump-probe delays, the simulated kinetic energy release differs significantly from the values obtained by a purely Coulombic approximation or a one-dimensional description of the dicationic potential energy surfaces, and shows a clear bifurcation near 12 fs, owing to non-adiabatic transitions through a conical intersection. The proposed approach is particularly suitable and efficient in simulating processes that highly rely on statistics or for identifying rare reaction channels. 
} \\
\end{tabular}

 \end{@twocolumnfalse} \vspace{0.6cm}
]

\renewcommand*\rmdefault{bch}\normalfont\upshape
\rmfamily
\section*{}
\vspace{-1cm}


\footnotetext{\textit{$^{a}$~J. R. Macdonald Laboratory, Department of Physics, Kansas State University, Manhattan, KS 66506, USA. E-mail: yijueding@gmail.com}}








\section{Introduction}

Photochemical reactions are vital and ubiquitous in our daily lives. Representative examples include photosynthesis, human vitamin D formation and light-harvesting devices\cite{fahhad2015}. Photochemical reactions under low temperature also play an important role in quantum control and quantum information processing\cite{bohn2017,pan2022,ding2016,ding2017}. Following molecular structures in real time during photochemical reactions is vital to understand and control the reactions for specific purposes. To this end, ultrafast imaging techniques are developed to study reaction dynamics in a time-resolved manner. These techniques include attosecond transient absorption spectroscopy (ATAS)\cite{atas} and time-resolved photoelectron spectroscopy (TRPES)\cite{trpes}, which provides indirect information on nuclear geometries by measuring the electronic energy levels or the kinetic energy of photoelectrons with extremely high temporal resolution. Other techniques such as ultrafast electron diffraction (UED)\cite{ued} and Coulomb explosion imaging (CEI)\cite{vager1989} provides direct information of the nuclear geometry by interpreting the diffraction pattern or measuring the momenta of ionic fragments. 

In particular, CEI is a powerful technique that is equally sensitive to both light and heavy atoms. CEI retrieves molecular structure by rapidly removing multiple electrons and measuring the asymptotic momenta of ionic fragments. Thanks to decades of development, CEI can be applied to molecules consisting of a couple of atoms to over ten atoms in a molecule\cite{reviewcei,martin2013,boll2022,li2022,lam2024}. In a time-resolved manner, CEI has been used to identify isomerization processes\cite{burt2018}, isolate a rare roaming channel\cite{endo2020}, or study the Jahn-Teller effect\cite{li2021,wang2024}.

Methyl iodide ($\text{CH}_3\text{I}$) is a prototype molecule to study photochemical reaction dynamics due to its simplicity and rich variety of photochemical phenomena, and thereby has attracted research interests for a long time\cite{mulliken,kent1972,shapiro2008,godwin1987,parker1998,corrales2014}.
CEI experiments have been conducted on this molecule both near the ground state equilibrium\cite{corrales2012,dingdajun2017} and during photodissociation in a time-resolved manner\cite{amini2018,felix2018,corrales2019,daniel2023}.
One intriguing feature is the nondiabatic dynamics during photodissociation 
owing to a conical intersection (CI) induced by the strong spin-orbit coupling effect. 
Non-adiabtic transitions near such a CI result in two dissociation channels corresponding to $\text{I}(^2\text{P}_{3/2})$ and $\text{I}^*(^2\text{P}_{1/2})$ products, respectively. 

In the theoretical side, several trajectory calculations have been performed on potential energy surfaces (PES) of $\text{CH}_3\text{I}$ excited states over time\cite{amatatsu1991,amatatsu1996,alekseyev2007}, as well as ab initio molecular dynamics (AIMD) calculations\cite{kamiya2019,wangch3i} and quantum mechanical calculations\cite{guo1992,xie2000,evenhuis2011} that have been conducted to study the A-band photodissociation. However, accurate simulations of both the photodissociation and the Coulomb explosion
process to quantitatively simulate the observables from
the pump-probe experiments have not been performed so far. Particularly, no molecular dynamics simulations have been conducted for the doubly charged ionic state, which is important for the Coulomb explosion process. 

In this work, we study the UV-induced dissociation of $\text{CH}_3\text{I}$ probed by CEI in a time-resolved manner through trajectory calculations on ab initio PESs of both the neutral molecule and the doubly charged cation.  
Since CEI is a statistical method that relies on a large number of events, theoretical methods like AIMD, which calculate electronic structures on-the-fly, may not be cost effective to run for a sufficient number of trajectories to obtain proper statistics.
To address this issue, we develop an efficient program that performs fast trajectory calculations on both the neutral and ionic surfaces with the inclusion of non-adiabatic effects. 

\section{Theory and computational methods}
\subsection{Ab initio electronic structure calculation}

All electronic structures are computed at the multi-reference configuration interaction (MRCI)\cite{mrci1,mrci2} level of theory. The active space is 6 electrons in 5 orbitals for both the neutral $\text{CH}_3\text{I}$ molecule and the $\text{CH}_3\text{I}^{2+}$ cation. The active space includes two nearly degenerate lone pair ($n$) orbitals of the I atom, the C-I bonding ($\sigma$) and antibonding ($\sigma^*$) orbitals, for the purpose of accurately describing the A-band excited states that are dominated by $n\rightarrow\sigma^*$ transitions. 
Besides these four orbitals, one nearly empty orbital above the $\sigma^*$ orbital is added to the active space of $\text{CH}_3\text{I}$ and one nearly doubly-occupied orbital below the $\sigma$ orbital is added to the active space of $\text{CH}_3\text{I}^{2+}$ to achieve better convergence of energy.

The state-averaged multi-configuration self-consistent field (SA-MCSCF)\cite{mcscf1,mcscf2,mcscf3} calculation is first carried out to obtain the reference wavefunctions. For the neutral $\text{CH}_3\text{I}$ molecule, 3 lowest singlet states and 3 lowest triplet states are computed and optimized with equal weights in the SA-MCSCF procedure to cover all A-band excited states. For the $\text{CH}_3\text{I}^{2+}$ cation, 7 singlets and 4 triplets are computed and optimized with equal weights in the SA-MCSCF procedure to cover all possible atomic terms of the $\text{I}^+$ cation. The MRCI calculation is then performed by adding single and double excitations to external configurations from the MCSCF reference wavefunctions. 

The spin-orbit (SO) coupling effect is treated using a state-interaction (SI) method. The SO eigen-energies are calculated by diagonalizing the SO coupled Hamiltonian, $\hat{H}_{el}+\hat{H}_{SO}$, under the basis of MCSCF wave functions, with the diagonal elements replaced by MRCI energies. 

The Dunning type correlation-consistent basis sets\cite{dunningbasis} are employed in our electronic structure calculations. Specifically, we use cc-pVTZ basis sets for hydrogen and carbon atoms. We also use cc-pVTZ-PP basis set for the iodine atom, where the inner 28 core electrons are described by a relativistic pseudo-potential\cite{pseudopotential,pseudopotential2}.

All ab initio calculations are performed using the MOLPRO quantum chemistry package\cite{molpro,molpro2}. 

\subsection{Analytic potential energy surfaces}

In order to obtain an accurate description of the electronic energies and the corresponding gradients, we calculate about 40,000 geometries for $\text{CH}_3\text{I}$ and about 20,000 geometries for $\text{CH}_3\text{I}^{2+}$. The PES is constructed under $\{R, \alpha,\theta\}$ degrees of freedom, where R is the C-I distance, $\alpha$ is the angle between the C-H vector and the $C_3$ axis of the $\text{CH}_3$ component, and $\theta$ is the angle between the C-I vector and the $C_3$ axis of the $\text{CH}_3$ component. 
The umbrella motion (characterized by $\alpha$) and the C-I bending (characterized by $\theta$) are strongly coupled with the C-I bond stretching during dissociation. The C-I bending motion also plays an important role in non-adiabatic transitions. 
Particularly, ab initio electronic structure calculations are performed on a very dense grid ($\Delta R=0.01 \text{\AA}$) near the conical intersection between $^3\text{Q}_0$ and $^1\text{Q}_1$ states in order to accurately characterize the nonadiabatic effects. 
Analytic representations of the adiabatic PESs are generated using a multivariate cubic spline interpolation algorithm.
Because the grid is sufficiently dense, such a direct interpolation is able to characterize the conical intersection with high fidelity. 

The ab initio potential energies are calculated up to $R_c=5$ $\text{\AA}$ for $\text{CH}_3\text{I}$ and up to $R^\mathrm{ion}_c=10$ $\text{\AA}$ for $\text{CH}_3\text{I}^{2+}$, beyond which the energy gradient of $\text{CH}_3\text{I}$ is small($|d V_{\text{CH}_3\text{I}}/d R|<0.01$ eV/\AA), and the potential of $\text{CH}_3\text{I}^{2+}$ is almost purely Coulombic($|d(V_{\text{CH}_3\text{I}^{2+}}-1/R)/d R|<0.01$ eV/\AA). In order to simulate the the pump-probe process at large delays (up to 500 fs) and to obtain the asymptotic momenta of the ionic fragments after Coulomb explosion, it is natural to extend the PES to infinite C-I distance.

The interaction between the neutral I and $\text{CH}_3$ fragments is dominated by the van der Waals force at large internuclear distances, which is in the order of $R^{-6}$.
We neglect this high-order interaction, and therefore the $\text{CH}_3\text{I}$ potential beyond $R_c$ is written as
\begin{equation}
  V_{\text{CH}_3\text{I}}\approx V_{\text{CH}_3}+V_\text{I} \qquad (R>R_c),
\end{equation}
where $V_{\text{CH}_3}$ only depends on the internal degrees of freedom of the $\text{CH}_3$ fragment, and $V_\text{I}$ is the internal energy of the I atom, which also defines the thresholds of the different reaction channels.

The interaction between the ionic $\text{I}^+$ and $\text{CH}_3^+$ fragments is dominated by the Coulomb force at large internuclear distances. 
We also include a $1/R^4$ term to describe the charge-dipole interaction\cite{chargedipole}.
This isotropic term characterizes both the charge-permanent dipole interaction after averaging the dipole orientation and the charge-induced dipole interaction. 
Thus, the $\text{CH}_3\text{I}^{2+}$ potential beyond $R^\mathrm{ion}_c$ is written as 
\begin{equation}
  V_{\text{CH}_3\text{I}^{2+}}\approx V_{\text{CH}_3^+}+V_{\text{I}^+}+(\frac{1}{R}-\frac{C_4}{R^4}) \qquad (R>R^\mathrm{ion}_c),
\end{equation}
where the coefficient $C_4$ is obtained by fitting the long range tails of the calculated potential energies.

\subsection{Photodissociation-Coulomb explosion trajectory calculation}
In order to directly simulate the observables from time-resolved CEI experiments, we develop a molecular dynamics program to simulate photodissociation followed by Coulomb explosion in a single trajectory calculation on pre-built analytic PESs of $\text{CH}_3\text{I}$ and $\text{CH}_3\text{I}^{2+}$. 

At the first step, the initial conditions are sampled from a classical Boltzmann distribution\footnote[2]{Zero-point energy (ZPE) is not included in this sampling method, so there is no ZPE leakage issue in the trajectory calculation. The ZPE leakage issue would not be significant even if ZPE were considered since the high-frequency CH stretching modes are absent due to reduced dimensionality. The agreement between our simulation and the experiment also justifies our assumption.} of normal mode coordinates of the ground electronic state at effective temperature $T=100\sim 200$ K. 
Since the PESs are constructed in reduced dimension, 
the normal mode coordinates $\eta_i(i=1,2,3)$ can be written as $\{\eta_1,\eta_2,\eta_3\}=\textbf{U}\{R,\alpha,\theta\}$, where $\textbf{U}$ is the transformation matrix. The Boltzmann probability distribution is given by 
\begin{equation}
    P(\eta_i,\dot{\eta}_i)\propto \exp \{-(m_i\dot{\eta}_i^2+m_i\omega_i^2\eta_i^2)/2k_BT\}\quad (i=1,2,3)
\end{equation}
where $k_B$ is the Boltzmann constant, $m_i$ is the generalized mass for each normal mode coordinate, and $\omega_i$ is the corresponding normal mode frequency. The normal mode analysis on the ground state PES reveals $\{\omega_1,\omega_2,\omega_3\}=\{530,880,1308\}\text{ cm}^{-1}$, corresponding to C-I stretching, C-I bending, and $\text{CH}_3$ umbrella motion, respectively.
These frequencies are in overall agreement with the reported experimental values $\{533,882,1251\}\text{ cm}^{-1}$\cite{alanko1989,alanko1996,perrin2016}.
The remaining internal degrees of freedom are kept the same as in the equilibrium geometry. 

For UV photon absorption near 266 nm, transition to the $^3\text{Q}_0$ state (mostly parallel) is much stronger than to other states in the A-band\cite{parker1998,alekseyev20072}. The momenta and positions of each atom do not change during the UV absorption (vertical transition). Thus, similar to other AIMD simulations performed on this molecule\cite{wangch3i,kamiya2019,corrales2019}, we start all trajectories on the $^3\text{Q}_0$ electronic state with initial conditions sampled on the ground electronic state. In our simulation, 
the equation of motion (EOM) is described by the Euler-Lagrange equation $\frac{d}{dt} \frac{\partial L}{\partial \dot{\textbf{q}}} -\frac{\partial L}{\partial \textbf{q}}=0$, where $\textbf{q}= \{\Psi,\Theta,\Phi,R,\alpha,\theta \}$. The Euler angles $\{\Psi,\Theta,\Phi\}$ are included in the EOM to describe the overall rotation of the entire molecule, which is likely to affect the rotational states of the resulting fragments after photodissociation and Coulomb explosion. 
The molecule is initially prepared at its rotational ground state with random orientation.
To propagate the trajectory, the EOM is integrated numerically using an adaptive Runge-Kutta algorithm with a desired relative error of $10^{-7}$. Non-adiabatic effects during the propagation are taken into account using a Landau-Zener surface hopping algorithm, which will be discussed in the following subsection. When the trajectory is calculated up to a pre-specified target time, the probe laser pulse is applied to doubly ionize the molecule and induces the Coulomb explosion.
We refer to this target time as the pump-probe delay. The pulse duration and its effects on the nuclear dynamics are not considered in our simulation. So the molecular dynamics are always conducted on field-free PESs. 
The photoionization event itself is assumed to occur instantaneously on the timescales considered in our simulation.
Therefore, the trajectory instantaneously hops to the $\text{CH}_3\text{I}^{2+}$ PES and continues to propagate until the Coulomb energy between ionic fragments is below 0.01 eV. To simplify the problem, non-adiabatic effects are not considered at the Coulomb explosion stage.
We call this type of trajectories photodissociation-Coulomb explosion (PD-CE) trajectories.

Since CEI is highly dependent on statistics, 
a large number of trajectories are required to obtain a statistically reliable interpretation of the CEI results. In our simulation, more than 40,000 PD-CE trajectories are calculated in total, specifically, about 20,000 in the asymptotic regime and about 20,000 for short pump-probe delays.

\subsection{Landau-Zener surface hopping algorithm}

In the present study, we employ a Landau-Zener model to treat the non-adiabatic effects in the trajectory calculation. The Landau-Zener surface hopping (LZSH) algorithm has been elaborated in previous studies, and was shown to yields reliable results\cite{domckelzsh,domckelzsh2,belyaevlzsh,lzshjctc}. The LZSH algorithm is easier to implement compared to the commonly used fewest-switch surface hopping (FSSH) algorithm\cite{tullyfssh}, while still yielding similar non-adiabatic results such as the time-dependent electronic population ratio\cite{domckelzsh,lzshjctc}.

The Landau-Zener non-adiabatic transition probability in the adiabatic picture is written as\cite{lebedev2011}
\begin{equation}
    P_{i\rightarrow j}=\text{exp}\left(-\frac{\pi}{2\hbar}\sqrt{\frac{Z_{ij}(\textbf{q}(t_0))^3}{\frac{d^2}{dt^2}Z_{ij}(\textbf{q}(t))|_{t=t_0} }}\right)
    \label{eq3}
\end{equation}
where $\textbf{q}(t)$ is the nuclear coordinate vector as a function of time, and $Z_{ij}=|V_i-V_j|$ is the absolute value of the potential energy difference between $i$ and $j$ adiabatic states. 
During the trajectory propagation, the value of $Z_{ij}$ 
is monitored at each time step. When a local minimum of $Z_{ij}$ as a function of time is attained, the non-adiabatic transition probability is calculated using Eq. \eqref{eq3}. Then, a random number is generated from a uniform distribution between 0 and 1, and is compared with the probability $P_{i\rightarrow j}$ to determine whether the trajectory should hop to a different adiabatic state. If a hop is confirmed, the velocity of the system is rescaled by a factor $\gamma$ to maintain the conservation of energy. This requirement can be written as
\begin{equation}
    \frac{1}{2}\dot{\textbf{q}}^T \textbf{M} \dot{\textbf{q}}+V_i(\textbf{q})=\frac{1}{2}\gamma\dot{\textbf{q}}^T \textbf{M} \gamma\dot{\textbf{q}}+V_j(\textbf{q}),
    \label{eq4}
\end{equation}
where $\dot{\textbf{q}}=d\textbf{q}/dt$ is the velocity vector of the nuclear coordinates, and $\textbf{M}$ is the mass matrix.

It should be noted that in this model a surface hopping decision can be made even when $Z_{ij}$ is large, but in our simulation all trajectories that undergo non-adiabatic transitions show $Z_{ij}< 0.2$ eV when the surface hoppings take place. Furthermore, non-adiabatic effects are only considered within $2.2 \text{ \AA} <R< 2.6 \text{ \AA}$.

\section{Results and discussions}
\subsection{Adiabatic potential energies of $\text{CH}_3\text{I}$ and $\text{CH}_3\text{I}^{2+}$}
\begin{figure}[h]
\centering
  \includegraphics[width=8.5cm]{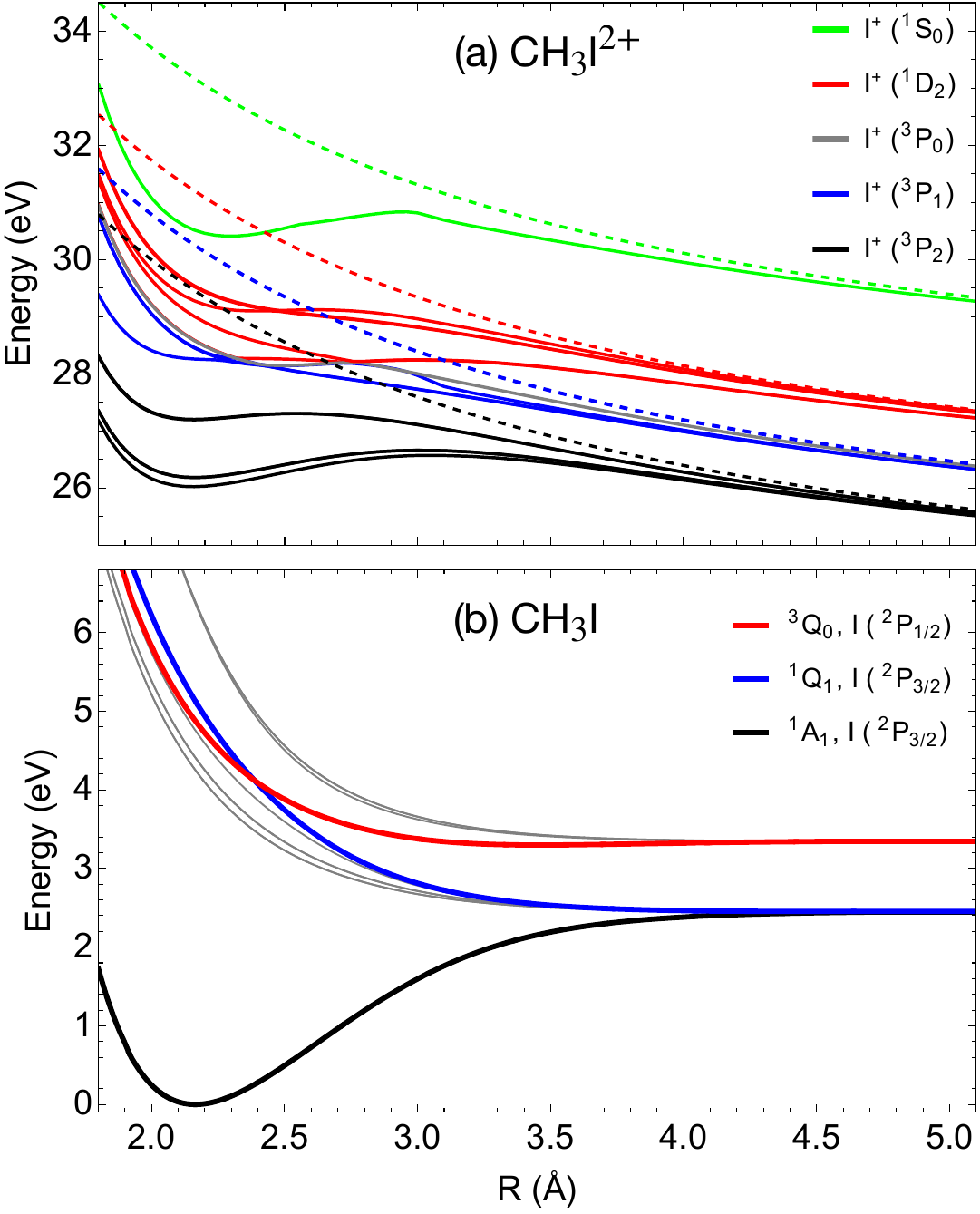}
  \caption{Ab initio potential energy curves for the $\text{CH}_3\text{I}^{2+}$ cation (a) and the neutral $\text{CH}_3\text{I}$ molecule (b) by stretching the C-I bond length $R$. All other degrees of freedom remain the same as in the equilibrium geometry. The energies are relative to the ground state energy of $\text{CH}_3\text{I}$ at equilibrium. In panel (a), solid curves are adiabatic potentials with different colors corresponding to different dissociation thresholds of the $\text{I}^+$ ions. Dashed curves are pure Coulomb potentials shifted to the corresponding $\text{I}^+$ thresholds for comparison. The potential curves in panel (b) correspond to the ground state and A-band excited states of the neutral molecule. The curves shown in thick black, blue, and red are the relevant states during UV induced photodissociation.}
  \label{fig1}
\end{figure}

The adiabatic PESs of $\text{CH}_3\text{I}$ and $\text{CH}_3\text{I}^{2+}$ are constructed based on high-level ab initio electronic structure calculations and multivariate cubic spline interpolations.
Figure \ref{fig1} shows the potential energies of $\text{CH}_3\text{I}$ and $\text{CH}_3\text{I}^{2+}$ by stretching the C-I bond length $R$. 
Fifteen ionic potential curves with thresholds corresponding to different internal states of $\text{I}^+$ are shown in Fig. \ref{fig1}(a), and are also compared with the pure Coulomb interactions between $\text{CH}_3^+$ and $\text{I}^+$. 
The energies of the ionic states at the Franck-Condon point correspond to the double ionization potentials of $\text{CH}_3\text{I}$. These energies range from about 26 eV to 31 eV in our calculation, in good agreement with previous theoretical and experimental studies\cite{markus2012,roos2017}.
The actual energies of all states are smaller than the pure Coulomb energies, owing to the attractive covalent interaction between $\text{CH}_3^+$ and $\text{I}^+$. 
This attractive interaction is particularly strong for small $R$. As the C-I distance increases, the potential becomes strongly repulsive and can be well approximated by pure Coulomb interactions for $R>4$ \AA.

Twelve neutral potential curves corresponding to the ground state and A-band excited states are shown in Fig. \ref{fig1}(b). At large C-I distance, these states dissociate to two atomic states of iodine, $\text{I}(^{2}\text{P}_{3/2})$ and $\text{I}^*(^{2}\text{P}_{1/2})$, with an energy splitting of 0.9 eV in our calculation.
Using the Mulliken notation\cite{mulliken}, the $^3\text{Q}_0$, $^1\text{Q}_1$ and $^3\text{Q}_1$ diabatic states are dipole allowed transitions from the ground state. 
An UV pulse near 260 nm mostly excites the molecule to the $^3\text{Q}_0$ state\cite{parker1998} through parallel transition. 
It is noted that a non-adiabatic crossing between $^3\text{Q}_0$ and $^1\text{Q}_1$ states is present at $R=2.38$ \AA, which will induce population transfer to the $^1\text{Q}_1$ state during photodissociation.

\begin{figure}[h]
\centering
  \includegraphics[width=8.5cm]{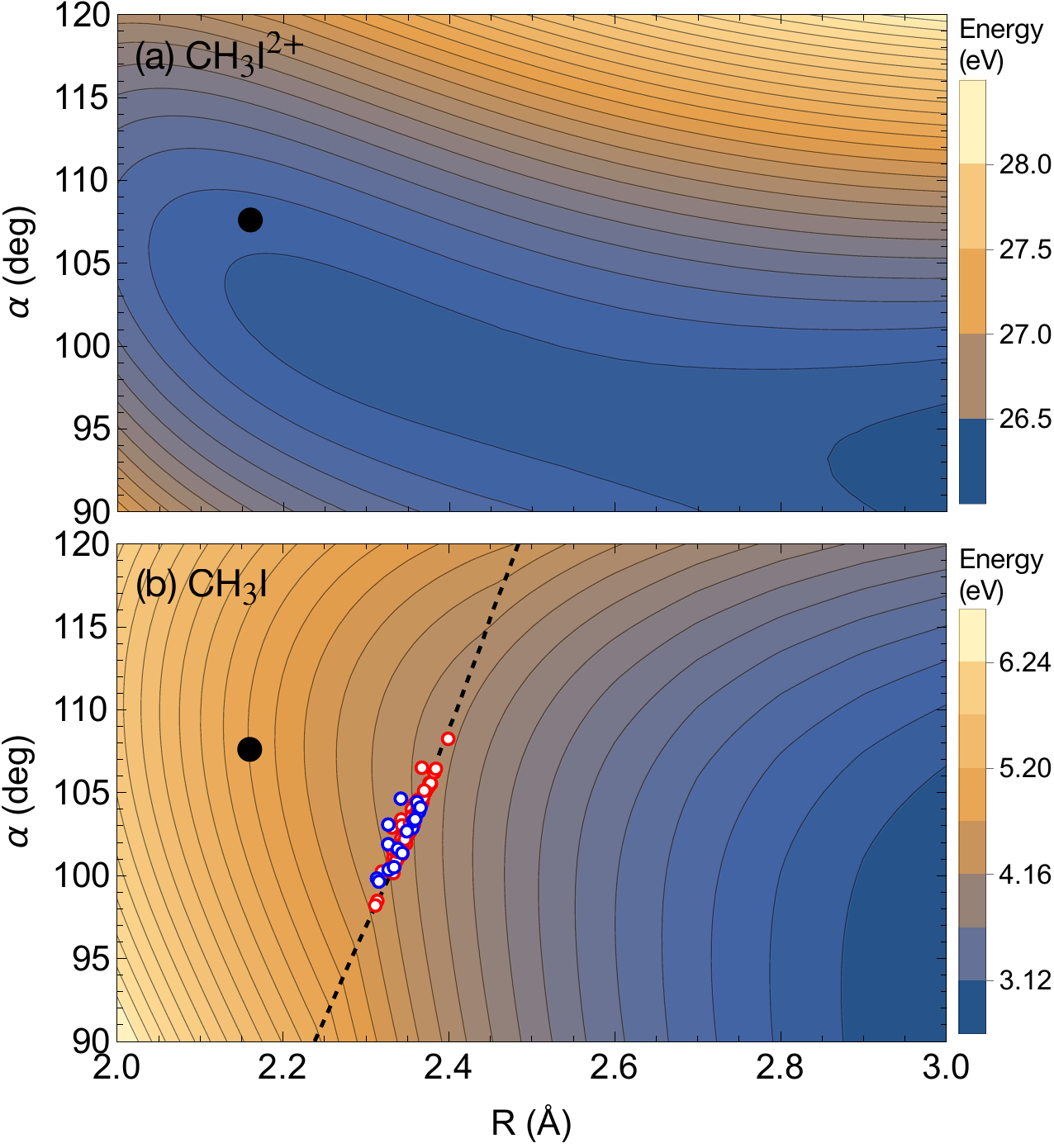}
  \caption{Potential energy surfaces of $\text{CH}_3\text{I}^{2+}$ and $\text{CH}_3\text{I}$ projected onto the $\alpha$ (angle between C-H vector and $C_3$ axis) and $R$ (C-I distance) plane. All other degrees of freedom are kept the same as in the equilibrium geometry such that the molecule(cation) is at $C_{3v}$ symmetry. The black dot denotes the equilibrium geometry. (a) The lowest adiabatic electronic state of $\text{CH}_3\text{I}^{2+}$ corresponding to the $\text{I}^+(^3\text{P}_2)$ dissociation threshold. (b) The 7th adiabatic electronic state of $\text{CH}_3\text{I}$ where all trajectory propagations start.
  The black dashed line represents the seam location where the energies of $^3\text{Q}_0$ and $^1\text{Q}_1$ states are degenerate, i.e. the conical intersections. The blue (hop) and red (no hop) circles indicate the geometries at which a surface hopping decision is made for 100 trajectories. 
  The energy difference between adjacent contours is 0.1 eV in Panel (a) and 0.14 eV in Panel (b).
  }
  \label{fig2}
\end{figure}

In previous CEI experiments, the measured delay-independent kinetic energy release (KER) show two major peaks centered around 4.4 eV and 5.2 eV for the doubly charged ion\cite{corrales2012,daniel2023}. Our simulation, as well as previous works\cite{corrales2012,daniel2023}, reveals that the KER peak near the 4.4 eV can be attributed to Coulomb explosions on the lowest ionic PES.\footnote[4]{The KER curves in Ref.\cite{corrales2012,daniel2023} are mislabeled. The labels associated to the $\text{I}^+(^3\text{P}_2)$ state and the $\text{I}^+(^1\text{D}_2)$ state should be swapped.}
Therefore, the molecule is highly likely ionized to the lowest $\text{CH}_3\text{I}^{2+}$ ionic state corresponding to $\text{I}^+(^3\text{P}_2)$ threshold after IR multi-photon absorption, and we simulate the Coulomb explosion part on this PES for most of our PD-CE trajectories. Moreover, conical intersections are absent on this PES, justifying our simplification of not considering non-adiabatic effects for the Coulomb explosion part. Figure \ref{fig2}(a) shows the lowest adiabatic PES of $\text{CH}_3\text{I}^{2+}$ as a function of the C-I distance $R$ and the angle $\alpha$ when the cation is in $C_{3v}$ symmetry. Contrary to the 1D potential curve shown in Fig.\ref{fig1}(a) and in other literature\cite{corrales2012,corrales2019}, there is no local minimum on this PES. 
Consequently, our simulation finds that no metastable $\text{CH}_3\text{I}^{2+}$ cations can be formed for this ionic state, which is in contrast to the results using 1D wave packet propagation reported in Ref. \cite{corrales2012}.
By further examination of all of the fifteen states in Fig. \ref{fig1}(a), only the state corresponding to the $\text{I}^+(^1\text{S}_0)$ threshold has a shallow well near $R=2.25$ \AA, which may support metastable $\text{CH}_3\text{I}^{2+}$ cations as will be discussed with more details in Section 3.4.

Since all trajectories propagate on adiabatic PESs, we also show the PES of the 7th adiabatic state of $\text{CH}_3\text{I}$ in Fig. \ref{fig2}(b). This adiabatic state is separated into $^3\text{Q}_0$ and $^1\text{Q}_1$ diabatic states by a "seam" line, which indicates the geometries of the conical intersections. All PD-CE trajectories are initially prepared on this adiabatic PES near the Franck-Condon (FC) point, located at $R_0=2.16$ $\text{\AA}$ and $\alpha_0=107.6$\textdegree. As shown in Fig. 2(b), when the trajectories propagate close to the seam line, non-adiabatic transitions are likely to take place according to the Landau-Zener model and the trajectories may hop to a different adiabatic state. 

\subsection{Simulating the kinetic energy release}
In a CEI experiment, the asymptotic momentum of each ionic fragment is measured via either velocity map imaging (VMI)\cite{burt2018,corrales2012,dingdajun2017,amini2018,felix2018,corrales2019} or COLd Target Recoil-Ion Momentum Spectrometer (COLTRIMS)\cite{vager1989,martin2013,boll2022,li2022,endo2020,daniel2023}. In the latter case, the measured kinetic energy release (KER) is the sum of the kinetic energies of $\text{I}^+$ and $\text{CH}_3^+$ fragments detected in coincidence. In our simulation, this corresponds to the translational kinetic energy of the relative motion between I($\text{I}^+$) and $\text{CH}_3$($\text{CH}_3^+$) fragments, which can be written as
\begin{equation}
 \mathrm{KER}=\mathrm{T_{trans}}=\frac{1}{2}\mu\dot{\textbf{R}}_{\mathrm{I-CH_3}}^2,
\end{equation}
where $\mu$ is the reduced mass between I($\text{I}^+$) and $\text{CH}_3$($\text{CH}_3^+$) fragments, $\textbf{R}_\mathrm{I-CH_3}$ is the displacement vector between I($\text{I}^+$) and the center of mass of $\text{CH}_3$($\text{CH}_3^+$), and $\dot{\textbf{R}}_\mathrm{I-CH_3}=d\textbf{R}_\mathrm{I-CH_3}/dt$ is the velocity of the relative motion.

The translational energy has contributions from two origins: the neutral excited state potential energy during photodissociation and the ionic state potential energy during Coulomb explosion. Thus, we further decompose the KER into two parts, 
\begin{equation}
    \mathrm{KER}=\mathrm{KER_{PD}}+\mathrm{KER_{CE}},
\end{equation}
where $\mathrm{KER_{PD}}$ and $\mathrm{KER_{CE}}$ are the kinetic energies gained from photodissociation and Coulomb explosion, respectively. 

\begin{figure}[h]
\centering
  \includegraphics[width=8.5cm]{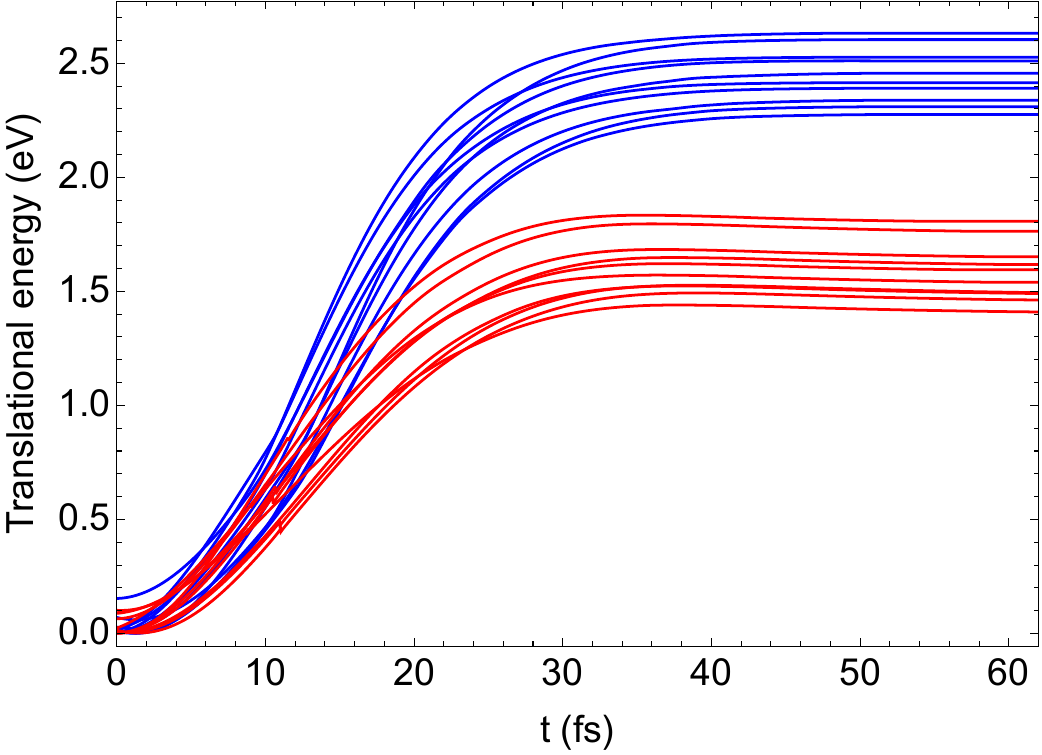}
  \caption{Translational kinetic energy of the relative motion between I and $\text{CH}_3$ fragments as a function of propagation time for 20 trajectories during photodissociation. The blue and red curves correspond to $\text{I}(^2\text{P}_{3/2})$ and $\text{I}^*(^2\text{P}_{1/2})$ dissociation channels respectively. 
  }
  \label{fig3}
\end{figure}

Figure \ref{fig3} shows the translational energy as a function of propagation time during photodissociation. This energy is also regarded as $\mathrm{KER_{PD}}$. During the initial 30 fs, the translational energy increases rapidly due to the repulsion of the excited state potential. Some trajectories (marked by red curves) show small jumps of the translational energy near 12 fs, which indicates that a surface hopping takes place and the kinetic energy is adjusted according to Eq. \eqref{eq4} to maintain the conservation of total energy. 
The translational energy saturates after about 50 fs for both $\text{I}$ and $\text{I}^*$ reaction channels, indicating the photodissociation is complete, and any further change in the final $\mathrm{KER}$ is attributed to Coulomb explosion contribution.
This photodissociation time scale is consistent with the calculation and measurement using XUV absorption spectroscopy \cite{wangch3i,changch3i}, in which one absorption peak disappears at about 50 fs because of the dipole forbidden I($^2\text{P}_{1/2}$) to I($^2\text{D}_{5/2}$) transition. 

When a one-dimensional (1D) model is applied to the PD-CE process, that is, considering the C-I distance $R$ as the only reaction coordinate and freezing all other degrees of freedom at the equilibrium geometry, $\mathrm{KER_{PD}}$ and $\mathrm{KER_{CE}}$ are approximated to be
\begin{equation}
\begin{split}
    \mathrm{KER_{PD}}(t)\approx V_{\text{CH}_3\text{I}}(R_0)-V_{\text{CH}_3\text{I}}(R(t)), \\
    \mathrm{KER_{CE}}(t)\approx V_{\text{CH}_3\text{I}^{2+}}(R(t))-V_{\text{CH}_3\text{I}^{2+}}(\infty), \\
    t=\int_{R_0}^R\frac{d\Tilde{R}}{\sqrt{2(V_{\text{CH}_3\text{I}}(R_0)-V_{\text{CH}_3\text{I}}(\Tilde{R}))/\mu}},
\end{split}
\label{eq7}
\end{equation}
where $t$ is the pump-probe delay. This 1D model is used for interpreting the experimental data in Ref.\cite{daniel2023}. At large pump-probe delays, the C-I distance $R$ is large and the photodissociation is complete. 
In that case, $\mathrm{KER_{PD}}\approx V_{\text{CH}_3\text{I}}(R_0)-V_{\text{CH}_3\text{I}}^{th}$ is approximately a constant value and $\mathrm{KER_{CE}}\approx 1/R(t)$ is approximately the pure Coulomb energy. 

\subsection{Coulomb explosion in the asymptotic region}

\begin{figure}[h]
\centering
  \includegraphics[width=8.5cm]{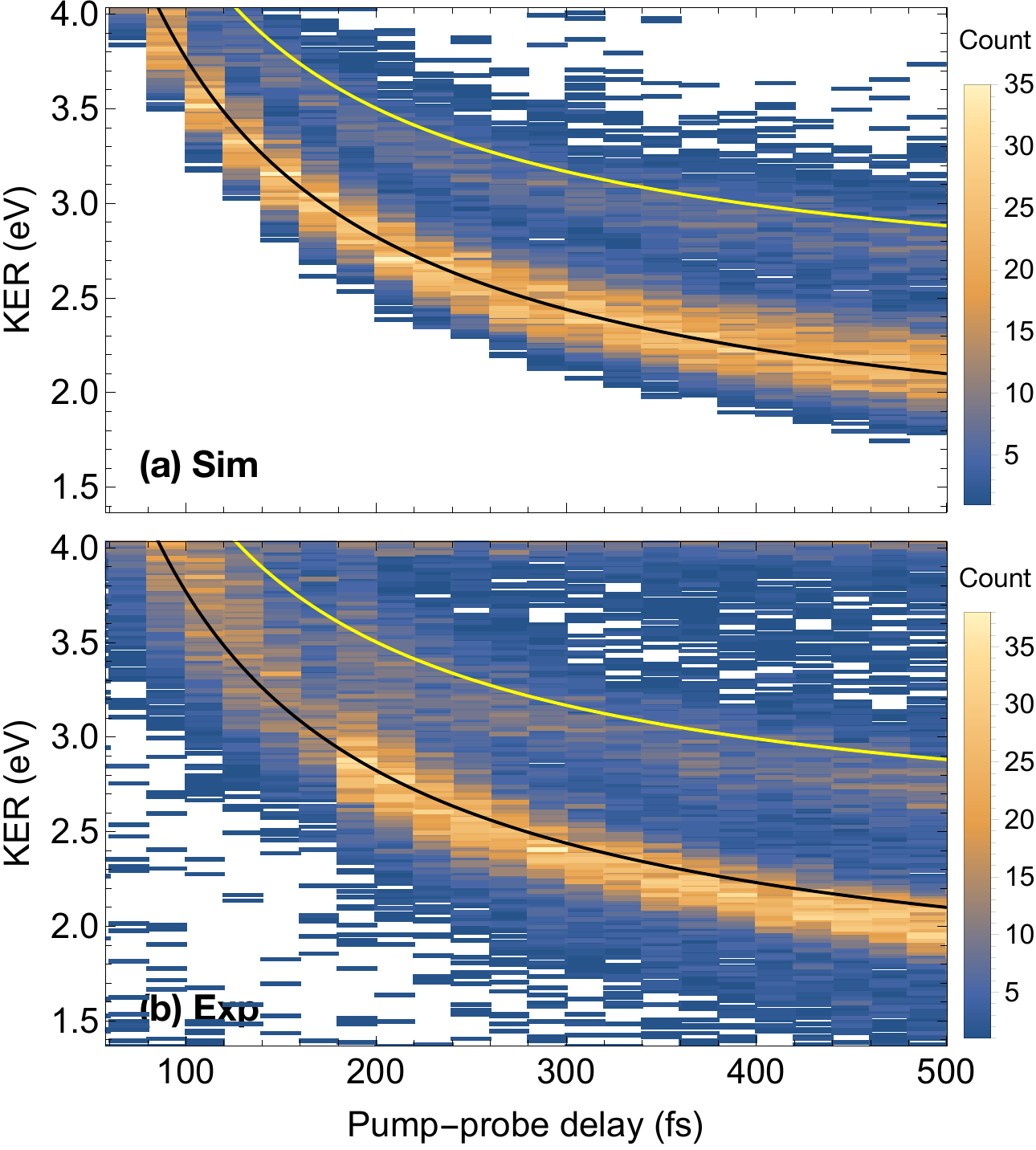}
  \caption{Kinetic energy release (KER) at different pump-probe delays ranging from 60 fs to 500 fs. (a) Signals from our simulation of about 20,000 PD-CE trajectories at $T=200$ K. The KER equals to the final translational kinetic energy of the relative motion between $\text{CH}_3^+$ and $\text{I}^+$ fragments. (b) Signals adapted from the experiment in Ref.\cite{daniel2023}. The KER is the sum of the kinetic energies of $\text{CH}_3^+$ and $\text{I}^+$ fragments measured in a coincident manner for a UV pump intensity of $1\times10^{13} \text{W}/\text{cm}^2$. The yellow and black curves are predictions using the 1D model in Eq.\eqref{eq7}. Signals in both (a) and (b) are counted using a bin size $h_t=20$ fs and $h_E=0.01$ eV.}
  \label{fig4}
\end{figure}

About 20,000 PD-CE trajectories are calculated at $T=200$ K with the pump-probe delay ranging from 50 fs to 500 fs and with a step of $\Delta t=1$ fs. Since previous studies have concluded that $\text{CH}_3\text{I}$ will be highly likely ionized to the lowest ionic states after IR multi-photon absorption\cite{corrales2012,dingdajun2017}, the Coulomb explosion part is simulated on the lowest PES corresponding to the $\text{I}^+(^3\text{P}_2)$ threshold. The resulting simulated delay-dependent KER signals are shown in Fig. \ref{fig4}(a), along with the relation between KER and the pump-probe delay predicted by the 1D model in Eq.\eqref{eq7}. The signals form two major bands, corresponding to the I and $\text{I}^*$ reaction channels. The lower band ($\text{I}^*$ channel) is significantly stronger than the upper band (I channel), indicating that a large fraction of trajectories undergo non-adiabatic transitions. The branching ratio $[\text{I}^*]/([\text{I}^*]+[\text{I}])$ is 0.82 in our simulation. This value is in overall agreement with simulations using the FSSH algorithm\cite{wangch3i,kamiya2019}, wave packet evolutions\cite{evenhuis2011} and experiments\cite{ch3iexp1,ch3iexp2,ch3iexp3,ch3iexp4}, which range from 0.65 to 0.85. 

The two bands show clear correlation between the KER and the pump-probe delay, which agree with the prediction of the 1D model. The energy of the upper band is slightly lower than that of the 1D model, because more potential energy is converted to the vibrational and rotational motion of $\text{CH}_3$ fragment in the $\text{I}^*$ reaction channel. In fact, over ninety percent of the potential energy is redistributed to the translational energy of $\text{CH}_3$ and I fragments during photodissociation, while a small fraction is redistributed to the vibrational and rotational motion of $\text{CH}_3$. Previous studies reveal that higher vibrational and rotational quantum states of $\text{CH}_3$ are excited in $\text{I}^*$ dissociation channel\cite{kamiya2019,guo1992}.

The simulated delay-dependent KER signals are also in good agreement with the experiment, as shown in Fig.\ref{fig4}(b), although the experiment shows slightly lower KER than the 1D model. We only compare our simulation with the experiment for KER below 4 eV because higher KER region in the experiment is dominated by delay-independent signals due to multi-photon ionization of ground-state $\text{CH}_3\text{I}$ by the probe pulse, which is beyond the focus of this work. A weak contribution at lower KER in the experiment is attributed to multi-photon Rydberg state excitation by the pump pulse, which is not accounted for in our simulation. Unlike the simulated signals that are evenly distributed in each time interval, more photodissociation signals are detected in the experiment at larger pump-probe delays, which leads to the intensity enhancement of the lower KER band as a function of time. The most likely reason for this is that the strong-field dissociative double ionization rate increases as a function of the internuclear distance $R$.
Such effects are well known, e.g. due to a varying ionization potential as a function of internuclear separation or due to resonance-like effects dubbed “enhanced ionization”\cite{bocharova2011,zuo1995}. 
Since we do not model the strong-field ionization process, such effects are absent in our simulation, and the ion pair yield in the simulation is, by definition, constant.

\begin{figure}[h]
\centering
  \includegraphics[width=8.5cm]{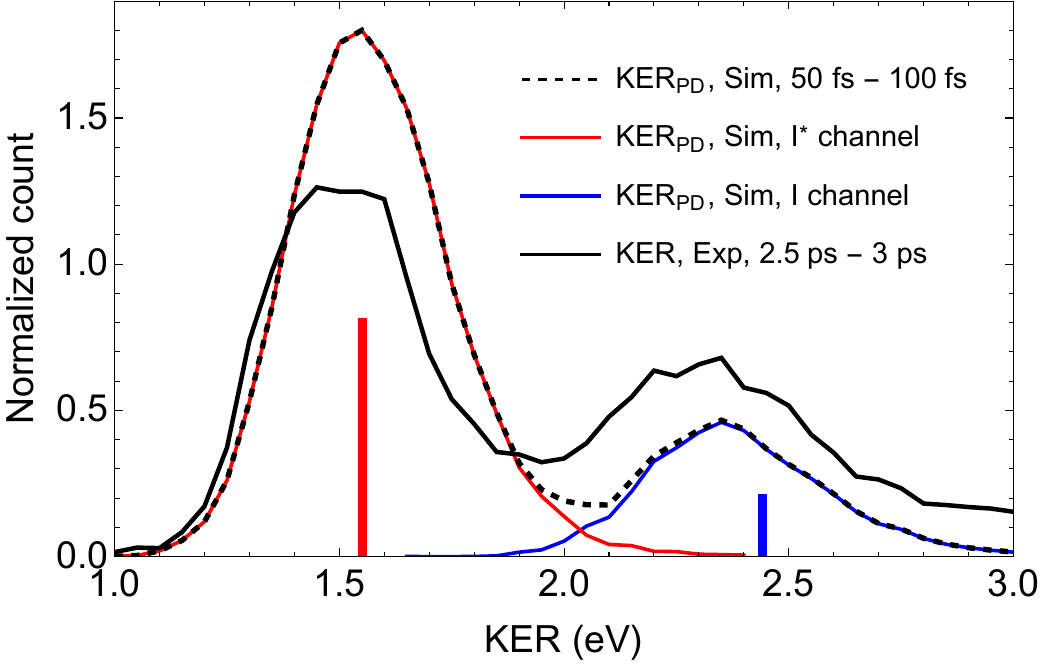}
  \caption{
  Experimental kinetic energy release (KER) signals (solid black, adapted from Ref.\cite{daniel2023}) compared with simulated $\mathrm{KER_{PD}}$ signals (dashed black). The experimental signals are integrated over the pump-probe delay from 2.5 ps to 3 ps for a UV pump intensity of $1\times10^{13} \text{W}/\text{cm}^2$. The simulated $\mathrm{KER_{PD}}$ include only the kinetic energy release during photodissociation, and are integrated over the propagation time from 50 fs to 100 fs. The blue and red curves are separate signals corresponding to $\text{I}$ and $\text{I}^*$ dissociation channels respectively in our simulation. The vertical lines indicate the potential energy differences between $^3\text{Q}_0$ state at the equilibrium geometry and the thresholds of $\text{I}^*$(red) and $\text{I}$(blue) channels, with the relative heights matching the $[\text{I}^*]/([\text{I}^*]+[\text{I}])$ branching ratio in our simulation. Signals are normalized by the total number of events between 1 eV and 3 eV.
  }
  \label{fig5}
\end{figure}

When the pump-probe delay is sufficiently large such that the Coulomb energy is negligible ($\mathrm{KER_{CE}}(t>2.5 \text{ ps})\approx 0$),
the measured KER is entirely due to the $\mathrm{KER_{PD}}$ during photodissociation. The $\mathrm{KER_{PD}}$ has already saturated at a propagation time $t=50$ fs ($\mathrm{KER_{PD}}(t>50 \text{ fs})\approx \text{Constant}$), as demonstrated in Fig. \ref{fig3}. Therefore, we can write the KER in this regime as follows: 
\begin{equation}
\begin{split}
\mathrm{KER}(t>2.5 \text{ ps}) & =\mathrm{KER_{CE}}(t>2.5 \text{ ps})+\mathrm{KER_{PD}}(t>2.5 \text{ ps}) \\
 &\approx \mathrm{KER_{PD}}(t>50 \text{ fs}) \\
 &\approx \text{Constant}
\end{split}
\label{eq8}
\end{equation}

Therefore, it is convenient to compare the simulated $\mathrm{KER_{PD}}$(t>50 fs) with the measured KER(t>2.5 ps), as shown in Fig. \ref{fig5}. The overall shape and peak positions of the simulated $\mathrm{KER_{PD}}$ signals shows good agreement with the experimentally measured KER. The relative height between the two peaks in our simulation is slightly different from the experiment, which reflects that the branching ratio in our simulation is higher than in the experiment. 
This can be attributed to the limitation of the present surface hopping model. Another reason is that we neglect the small fraction of dipole transitions to the $^1\text{Q}_1$ state that mostly leads to dissociations into the I channel.

\subsection{Coulomb explosion at short pump-probe delays}

\begin{figure*}
\centering
  \includegraphics[width=14cm]{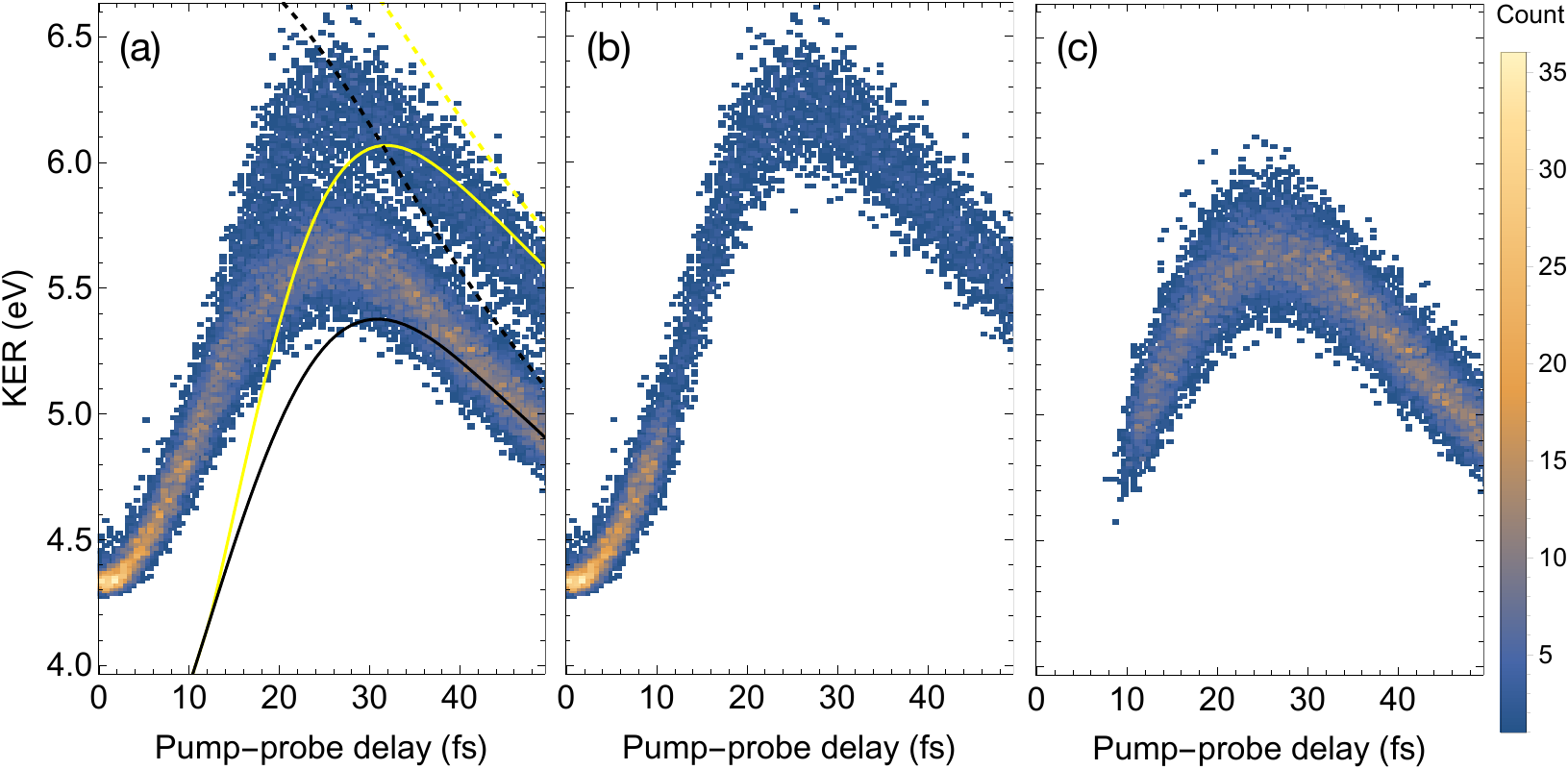}
  \caption{Simulated kinetic energy release at different pump-probe delays ranging from 0 fs to 50 fs using about 20,000 PD-CE trajectories at $T=100$ K. The Coulomb explosion is simulated on the lowest ionic potential energy surface of $\text{CH}_3\text{I}^{2+}$ which corresponds to $\text{I}^+(^3\text{P}_2)$ final dissociation limit. (a) Signals including all active states during photodissociation, which will lead to both the $\text{I}$ and $\text{I}^*$ channels. Results using the 1D model in Eq. \eqref{eq7} are shown in solid yellow and black curves. Results assuming pure Coulomb interaction are shown in dashed yellow and black curves. (b) Signals only from the 7th adiabatic state during photodissociation, which correspond to the  $\text{CH}_3+\text{I}$ channel. (c) Signals only from the 9th adiabatic state during photodissociation, which corresponds to the $\text{CH}_3+\text{I}^*$ channel at long distance. All signals are counted using a bin size $h_t=0.5$ fs and $h_E=0.01$ eV.}
  \label{fig6}
\end{figure*}

At short pump-probe delays (t<50 fs), the $\text{CH}_3\text{I}$ molecule is in the intermediate states of photodissociation.
Non-adiabatic transitions also take place at this time scale. To investigate the non-adiabatic dynamics in this regime,
additional 20,000 PD-CE trajectories are calculated with initial conditions sampled at $T=100$ K and with a pump-probe delay step $\Delta t=0.1$ fs up to 50 fs. 
We have used a lower temperature for sampling the initial conditions than in the simulations shown above in order to have less overlap between the signal from the two dissociating states.
The Coulomb explosion part is still simulated on the lowest PES corresponding to $\text{I}^+(^3\text{P}_2)$ threshold. The resulting delay-dependent KER is shown in Fig. \ref{fig6}. A bifurcation feature is present between 10 fs and 20 fs in Fig. \ref{fig6}(a), which indicates the occurrence of non-adiabatic transitions. The photodissociation separates to two channels leading to I and I* products. The simulation also shows that the 1D model discussed above is a poor approximation at short pump-probe delays, 
because the C-I stretching is coupled with the $\text{CH}_3$($\text{CH}_3^+$) umbrella motion and the C-I bending at small C-I distances, which is not characterized by the 1D model. 
Not only does the 1D model predict a fake potential well on the $\text{I}^+(^3\text{P}_2)$ ionic PES, but it also underestimates the KER. 
Fig. \ref{fig6}(b) shows that only a small fraction of trajectories (about 20 percent) undergo adiabatic passage and dissociate into the I channel. Most trajectories (about 80 percent) start to hop to the 9th adiabatic state near 12 fs and dissociate into the $\text{I}^*$ channel, as shown in Fig. \ref{fig6}(c). 

The bifurcation of KER signals has not been observed in CEI experiments so far. One reason is that the conical intersection is very close to the FC point and $\text{CH}_3$ is a very light fragment. Thus, non-adiabatic transitions occur within 20 fs after UV excitation, which requires pump and probe pulses with very high temporal resolutions to resolve. One option to alleviate this difficulty might be choosing a molecule with heavier fragments such as $\text{CF}_3\text{I}$. An alternative method to map the non-adiabatic transitions is using ATAS, which has been applied to methyl halide in previous works\cite{timmers2019,changch3i,wangch3i}. 

\begin{figure}[h]
\centering
  \includegraphics[width=8.5cm]{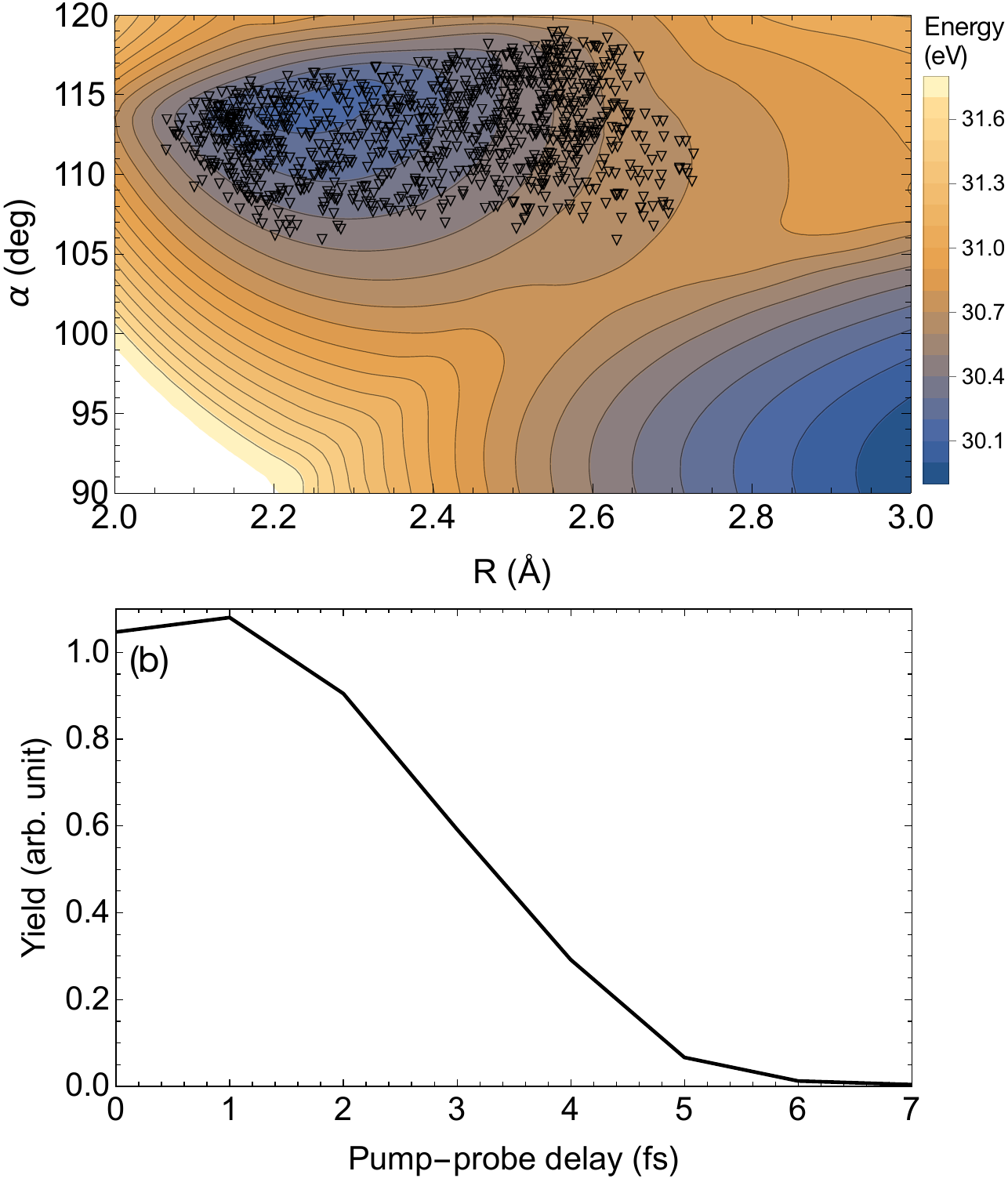}
  \caption{Creation of metastable $\text{CH}_3\text{I}^{2+}$ dications. (a) Potential energy surface of $\text{CH}_3\text{I}^{2+}$ corresponding to the $\text{I}^+(^1\text{S}_0)$ threshold, projected onto the $\alpha$ and $R$ plane. The energy difference between adjacent contours is 0.1 eV. 4,000 PD-CE trajectories are calculated at $T=100$ K with the CE dynamics occuring on this surface. The black triangles indicate the final geometries of 956 PD-CE trajectories that remain bound after propagating for over 2 ps on the ionic PES, and are thus counted as $\text{CH}_3\text{I}^{2+}$ signals. (b) Simulated $\text{CH}_3\text{I}^{2+}$ ion yield as a function of pump-probe delay.}
  \label{fig7}
\end{figure}

As mentioned before, only the ionic state associated with the $\text{I}^+(^1\text{S}_0)$ threshold has a shallow well on the PES and can support a metastable $\text{CH}_3\text{I}^{2+}$ state, as shown in Fig. \ref{fig7}(a). But it should be noted that this channel has a rather small transition probability during the IR multi-photon ionization. In order to investigate the origin of $\text{CH}_3\text{I}^{2+}$ signals observed in the CEI experiment\cite{daniel2023}, we calculate 4,000 PD-CE trajectories at very short pump-probe delays ($t<10$ fs), for which the Coulomb explosion is simulated on the $\text{I}^+(^1\text{S}_0)$ ionic PES. 956 trajectories form $\text{CH}_3\text{I}^{2+}$ states and do not break up into ionic fragments after propagating for over 2 ps, The final geometries are shown on top of the PES in Fig. \ref{fig7}(a). The $\text{CH}_3\text{I}^{2+}$ signal has a maximum yield near 1 fs, and decays rapidly with the pump-probe delay, as shown in Fig. \ref{fig7}(b). We therefore conclude that the $\text{CH}_3\text{I}^{2+}$ signal detected in the CEI experiment\cite{daniel2023} is due to this shallow well on the ionic PES, and the observed enhancement of the $\text{CH}_3\text{I}^{2+}$ yield near zero pump-probe delay is also consistent with the pathway described here, but it is broadened in the experiment due to  the cross correlation between the pump and the probe pulses. 
The internal states of $\text{CH}_3\text{I}^{2+}$ may be confirmed by photoelectron-photoelectron-photoion coincidence experiments.

\section{Conclusion and outlook}
In this work, we have employed an effective method to directly simulate the photodissociation followed by Coulomb explosion of methyl iodide with the inclusion of non-adiabatic effects. Potential energy surfaces of both the neutral molecule and the doubly charged cation are built upon high-level ab initio electronic structure calculations. More than 40,000 trajectories are calculated on both the neutral and ionic PESs to obtain a statistically reliable interpretation of the delay-dependent KER signals. The non-adiabatic effects are treated using a Landau-Zener surface hopping algorithm, which yields a branching ratio in overall agreement with previous theories and experiments. Our simulation of the delay-dependent KER at large pump-probe delays shows good agreement with a recent experiment using coincident ion momentum imaging techniques\cite{daniel2023}, with the KER signals reflecting two reaction channels. At short pump-probe delays, the simulation shows a clear bifurcation of the KER signals near 12 fs, indicating the occurrence of non-adiabatic transitions through a conical intersection between the $^3\text{Q}_0$ and $^1\text{Q}_1$ states. 
The bifurcation of the KER signals in our simulation is statistically reliable, and our simulation directly predicts experimental KER observables with good accuracy, which is beyond the scope of previous simulations using averaged AIMD trajectories\cite{corrales2019}.
Moreover, the simulation shows that metastable $\text{CH}_3\text{I}^{2+}$ ionic states can be formed with small probability on one of the high-lying ionic surfaces near zero pump-probe delay. 

This work reveals that dynamics on the ionic PES plays an important role when modeling experimental pump-probe observables such as the KER, especially in the regime of short pump-probe delays which cannot be described accurately by a simple 1D model. Our work also confirms that the Coulomb explosion imaging technique can effectively be used for direct visualization of conical intersections in coordinate space, provided sufficient temporal resolution. 

The computational procedure used in this work is efficient in calculating trajectories with non-adiabatic effects as we build PESs beforehand using a streamlined interpolation scheme, which can be applied to molecular dynamics simulations that highly rely on statistics, e.g., to identify reaction channels with small probabilities such as roaming. 
In the future, we plan to employ this approach to investigate photo-isomerization processes such as bond rearrangement in diiodomethane and ring opening in thiophenone. 


\section*{Conflicts of interest}
There are no conflicts to declare.

\section*{Acknowledgements}

YD thanks Sabre Kais for discussions at the early state of this study. YD thanks Panwang Zhou and Runze Liu for helpful advice on the ab initio calculations and non-adiabatic dynamics. 
The authors thank Farzaneh Ziaee for providing the experimental data shown in Figures 4 and 5. The authors also thank B. D. Esry for suggestions on improving the manuscript.
This work is supported by the Chemical Sciences, Geosciences, and Biosciences Division, Office of Basic Energy Sciences, Office of Science, U.S. Department of Energy, Grant no. DE-FG02-86ER13491.


\balance


\bibliography{rsc} 
\bibliographystyle{rsc} 

\end{document}